\documentclass[12pt,a4paper]{amsart}
\usepackage[utf8]{inputenc}

\usepackage[sort,round]{natbib}
\usepackage{doi}

\usepackage[left=1in,right=1in,top=1in,bottom=1in]{geometry}

\usepackage{multirow}
\usepackage{amsmath}
\usepackage{xfrac}
\usepackage{amsfonts}
\usepackage{amssymb}
\usepackage{graphicx}
\usepackage{float}
\usepackage{hyperref}
\usepackage{bookmark}

\newcommand{\der}[2]{\frac{\mathrm{d} #1}{\mathrm{d} #2}}

\usepackage{subcaption}
\usepackage{chngcntr}
\counterwithin{figure}{section}
\usepackage{color}

\title{A holed membrane at finite equibiaxial stretch}
\author[Friedberg and deBotton, Ben-Gurion Univ.]{{Idan Z. Friedberg and Gal deBotton}\\
{\footnotesize{\sl Dept.~of Mechanical Engineering, Ben-Gurion University,} \\
{\footnotesize {\sl Beer-Sheva 8410501, Israel}}}\\
}

\def \o{\mathrm{o}}
\def \i{\mathrm{i}}

\begin{document}
\maketitle

\begin{abstract}
The deformation and stress distribution in a stretched thin neo-Hookean circular membrane with a hole at its center are analyzed within the framework of finite deformation elasticity.
Initially, we derive a simple form for the differential governing equation to the problem.
This enables us to introduce a closed-form solution in the limit of infinite stretch.
Subsequently, we propose approximate solutions for intermediate and large deformations.
These approximations approach the exact solutions in the limits of small and infinite stretches.
The transition stretch at which the membrane behavior switches from the intermediate to the large deformation approximation is determined too.
Comparison of our solution and approximations to corresponding numerical results reveal a neat agreement for any stretch and ratio between the hole to the membrane radii.

In the limit of large stretches and a small hole, the ratio of the hoop stress at the hole boundary to the nominal stress is 4, 
which is twice the corresponding ratio in the small deformation limit.
Comparison of the strain energy stored in the membrane to the one in a membrane without a hole reveals that only at finite stretches the difference between these energies becomes meaningful.
This implies that it is likely that a flaw in a membrane will tear out only at a finite level of stretches.
\end{abstract}

\setcounter{page}{1}
\pagenumbering{arabic}

\newdimen\origiwspc
\origiwspc=\fontdimen2\origibaselineskip
\origibaselineskip=\baselineskip

\section{Introduction}
The problem of an equibiaxially stretched membrane with a hole at its center is a frequently encounter one.
The well known axisymmetric solution in the limit of infinitesimal deformations (e.g.  \cite{Shames1992:1}) is crucial for analyzing failure due to stress concentrations around the hole.
Moreover, the asymptotic solution in the limit a small hole is frequently used for estimating the stress in many other problems containing small holes.
However, the known solution for infinitesimal deformations is not applicable for materials such as elastomers and tissues that undergo finite deformations.
For these stretchable materials this problem need to be analyzed within the framework of \emph{finite deformation elasticity} \cite{OgdenR.W.1997Ned}.

Previous analyses \citep[e.g.,][]{rivlin1951large, yang1967stress, wong1969large}
dealt with a general Mooney-Rivlin material \citep{mooney1940, rivlin1948}, with the neo-Hookean material as a special case, and a wide range boundary conditions. Exact analytical results were developed in the limit of small deformations. 
Approximated expressions based on expansion series about these exact solution were proposed and compared with corresponding numerical results for a narrow range of ratios between the hole and the membrane radii.
\citet{yang1967stress} and \citet{wong1969large} further pointed out how their approximations break down in the case of traction free boundary condition at the hole.
\citet{haughton1991exact} solved a similar plane-stress problem for the \cite{varga1966stress} material, revealing that for this material the membrane thickness remains uniform.
Various numerical analyses of similar problems and the related cavitation problems were conducted with different hyperelastic materials  \citep[e.g.,][]{haughton1990cavitation, cohen2010cavitation, sang2015}.

Herein, we tackle the problem of a finitely stretched incompressible neo-Hookean circular membrane with a traction free hole of an arbitrary size at its center.
Initially, we introduce a simple non-dimensional, non-linear second order ordinary differential equation governing the boundary value problem.
This further leads to a closed-form solution for infinite stretching of the membrane and approximations for intermediate and large deformations.
The approximations are compared to corresponding numerical solutions of the problem, confirming their accuracy.
Additionally, the strain energy and stress distributions are analyzed and compared to the nominal cases of a membrane without a hole and a circumferentially stretched thin ring.
The former comparison implies that a small flaw in stretched membrane is likely to tear out only at finite strains.

\section{Background}
The deformation of a 3-dimensional body 
from a reference (undeformed) configuration $\mathcal{B}_\mathrm{r}\subset \mathbb{R}^3$ to a current (deformed) configuration $\mathcal{B}\subset \mathbb{R}^3$,
can be described by the bijection mapping of each material point $P$ at a reference position $\mathbf{X}\in \mathcal{B}_\mathrm{r}$ to its corresponding current position $\mathbf{x} =  \boldsymbol{\chi}(\mathbf{X})\in \mathcal{B}$.
Both configurations can be represented in a coordinate system.
We denote the unit vectors along chosen referential and current coordinates ($\alpha$ and $i$ respectively) as $\mathbf{\hat{E}}_\alpha$ and $\mathbf{\hat{e}}_i$ respectively.

The deformation gradient is
\begin{equation}
 \mathbf{F}\equiv
 \mathrm{Grad}(\mathbf{x}),
\end{equation}
where
$\mathrm{Grad}$ is the gradient with respect to the reference position $\mathbf{X}$.
The change in the volume of a material element is
\begin{equation}
J \equiv \det(\mathbf{F}) = \der{v}{V},
\end{equation}
where $\mathrm{d}V$ is its referential volume and $\mathrm{d}v$ its current volume.
The right and left \emph{Cauchy-Green deformation tensors} are 
\begin{equation}
\mathbf{C}=\mathbf{F}^T\mathbf{F}\qquad
\text{and}\qquad 
\mathbf{b}=\mathbf{F}\mathbf{F}^T,
\end{equation}
respectively.

If
no body forces are at play, the equilibrium equation for linear momentum
\begin{equation}\label{current equilibrium1}
\mathrm{div}(\boldsymbol{\sigma})= \mathbf{0},
\end{equation}
where $\boldsymbol{\sigma}$ is the \emph{Cauchy} stress (or true stress) and $\mathrm{div}$ is the divergence with respect to the current position.
The balance of angular momentum implies that $\boldsymbol{\sigma}$ is symmetric.
Using \emph{Nanson's formula} for the transformation of area elements, the force on a current area element $\mathrm{d}\mathbf{s}$ is equated to the traction on a referential area element $\mathrm{d}\mathbf{S}$,
\begin{equation}
\boldsymbol{\sigma}^T \mathrm{d}\mathbf{s} = J\boldsymbol{\sigma}^T\mathbf{F}^{-T} \mathrm{d}\mathbf{S} \equiv \mathbf{P}\mathrm{d}\mathbf{S},
\end{equation}
where $\mathbf{P}$ is defined as the \textit{Piola} stress (or nominal stress). Substituting $\boldsymbol{\sigma} = J^{-1}\mathbf{F}\mathbf{P}^T$ into 
\eqref{current equilibrium1}
yields the referential equilibrium equation
\begin{equation}
\mathrm{Div}(\mathbf{P})=\mathbf{0}, \label{referencial equilibrium}
\end{equation}
where $\mathrm{Div}$ is the divergence with respect to the reference position.

The constitutive behavior of hyperelastic materials can be expressed in terms of a \emph{strain energy-density function} (SEDF) $W$ such that
\begin{equation}
P_{i \alpha} = \frac{\partial W}{\partial F_{i \alpha}}.
\end{equation}
A simple constitutive relation that describes the behavior of many materials in the small to intermediate deformation range is the \emph{incompressible neo-Hookean} material,
for which
\begin{equation}
W =\frac{\mu}{2} (I_1-3) \qquad \text{and} \qquad I_3 \equiv 1,
\end{equation}
where $I_1 = \mathrm{tr}(\mathbf{C})$ and $I_3 = \det (\mathbf{C}) = J^2$.
Accordingly,
\begin{equation}
\mathbf{P} = \mu \mathbf{F}-p\mathbf{F}^{-T}\qquad
\text{and} \qquad
\boldsymbol{\sigma} = \mu \mathbf{b}-  p\mathbf{1}, \label{constitutive}
\end{equation}
where $\mathbf{1}$ is the identity tensor and $p$ is a Lagrange multiplier that represents a pressure like term.

\section{Analysis}
\begin{figure}[t]
 \centering
 \includegraphics[width = \textwidth]{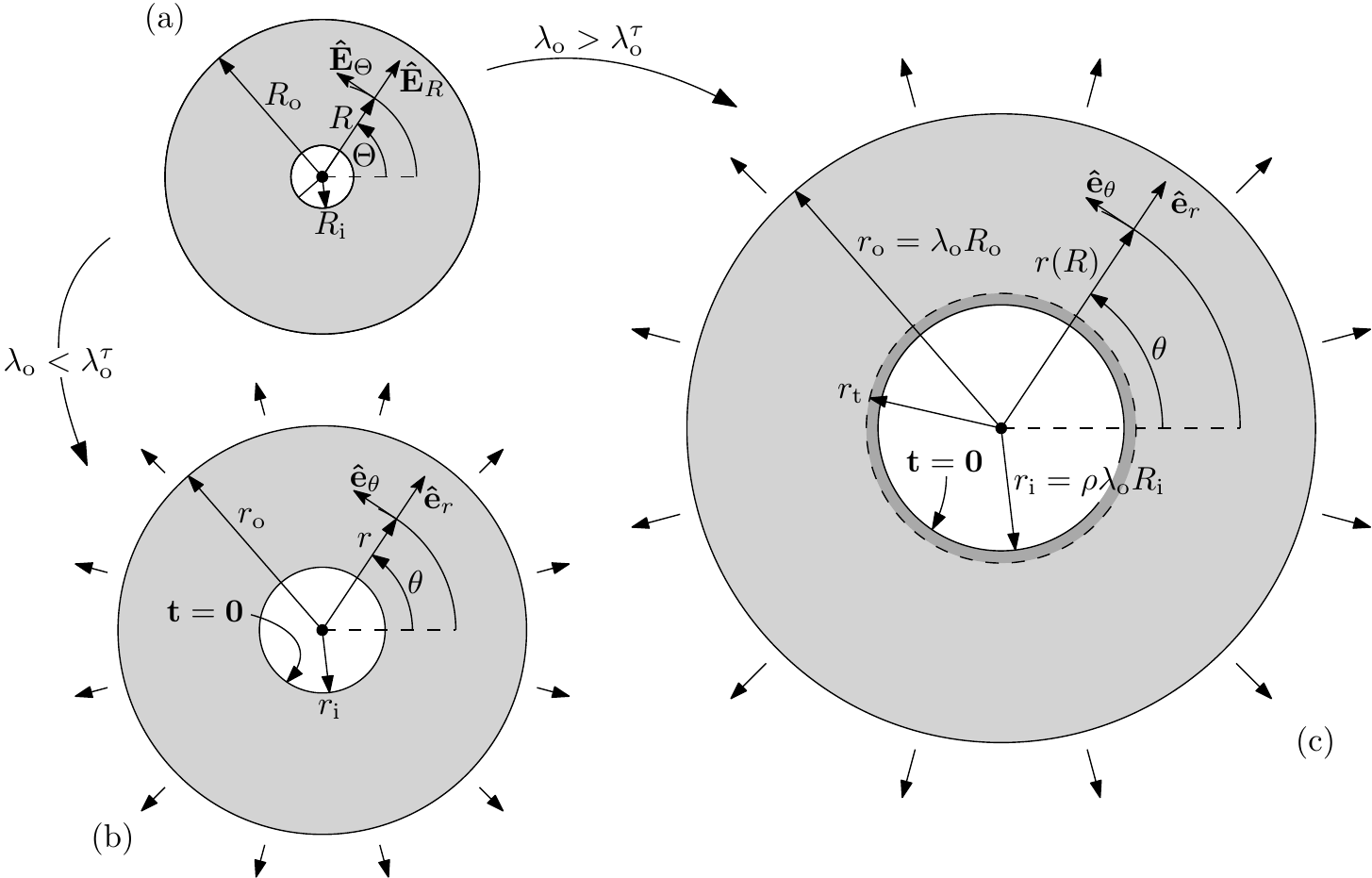}
 \caption{\small{A biaxially stretched punctured membrane with $R_\i = 0.2 R_\o$ in its
 (a) reference configuration, (b) moderately deformed configuration with $\lambda_\o = 1.3<\lambda_\o^\tau$ 
 and (c) severely deformed configuration $\lambda_\o = 2.0>\lambda_\o^\tau$. 
 The transition radius $r_\mathrm{t}$ between the inner and the outer regions of the severely deformed membrane is depicted in (c).}}
 \label{drawings}
\end{figure}

Consider a thin circular incompressible neo-Hookean membrane with a hole at its center (see Fig. \ref{drawings}a).
In its referential state the outer radius of the membrane is $R_\mathrm{o}$ and the radius of the hole is $R_\mathrm{i}$.
The membrane is subjected to biaxial stretch $\lambda_\mathrm{o}$ at the outer radius (see Fig. \ref{drawings}b and c) such that the outer radius in the deformed state is $r_\mathrm{o} = \lambda_\mathrm{o}R_\mathrm{o}$.
The circumference of the hole and the lateral faces of the membrane are stress free.
We recall that a well-known solution exists for the case where the punctured membrane is subjected to \emph{plane-strain} condition with the axial stretch being fixed \citep[e.g.,][]{OgdenR.W.1997Ned}. 

We examine this problem in polar coordinates $\{R, \Theta, Z\}$ in the reference configuration
and $\{r,\theta,z\}$ in the current configuration,
where the axial direction ($Z$ and $z$) is perpendicular to the membrane plane.
Since the problem is axisymmetric, for any point
$\mathbf{X}= R\mathbf{\hat{E}}_R + Z \mathbf{\hat{E}}_z$ in the reference configuration
we assume the mapping
\begin{equation}
\mathbf{x}= r(R)\mathbf{\hat{e}}_r + Z\zeta(R) \mathbf{\hat{e}}_z,
\end{equation}
where $\mathbf{\hat{e}}_r = \mathbf{\hat{E}}_R$ and $\mathbf{\hat{e}}_z = \mathbf{\hat{E}}_Z$ are the unit vectors of the polar system in the current and the reference configurations, respectively.

The boundary conditions at the outer and inner radii are
\begin{equation}\label{Boundary Conditions}
r(R_\mathrm{o})=\lambda_{\mathrm{o}}R_{\mathrm{o}},\ 
\sigma_{rr}(R_\mathrm{i}) = \sigma_{rz}(R_\mathrm{i}) = \sigma_{r\theta}(R_\mathrm{i}) = 0.
\end{equation}

Since lateral boundaries of the membrane are stress free, on account of the small thickness of the membrane we assume the \emph{plane stress condition} 
\begin{equation}\label{plane stress}
 \sigma_{zz} = \sigma_{rz} = \sigma_{\theta z} = 0,
\end{equation}
within the membrane.

The deformation gradient of the assumed mapping is
\begin{equation}
\mathbf{F}=\left[\begin{array}{ccc}
r_{,R} & 0 & 0\\
0 & \frac{r}{R} & 0\\
Z\zeta_{,R} & 0 & \zeta
\end{array}\right].
\end{equation}
We denote the radial, hoop and axial stretches
\begin{equation}
 \lambda_r \equiv  F_{rR} = r_{,R},\qquad
 \lambda_\theta \equiv F_{\theta \Theta} = \frac{r}{R}\qquad
 \text{and} \qquad
 \lambda_z \equiv F_{zZ} = \zeta,
\end{equation}
respectively.
Thanks to the negligible thickness of the membrane we set $F_{zR} = Z \zeta_{,R}=0$, and note that this is in agreement with the assumed plane stress condition.

Incompressibility yields that $J = r_{,R}\frac{r}{R}\zeta = 1$,
therefore the axial stretch is
\begin{equation}
\lambda_z = \zeta = r_{,R}^{-1}\frac{R}{r}.\label{zeta}
\end{equation}
Substituting $\mathbf{F}$ into the constitutive relations
\eqref{constitutive} leads to
\begin{gather}
{\boldsymbol{\sigma}}=
\mathrm{diag}\left\{ \mu r_{,R}^{2}-p,\mu\frac{r^{2}}{R^{2}}-p,\mu\zeta^{2}-p\right\},\\
{\mathbf{P}} = 
\mathrm{diag}\left\{ \mu r_{,R}-\frac{p}{r_{,R}},\mu\frac{r}{R}-\frac{pR}{r},\mu\zeta-\frac{p}{\zeta}\right\}.
\end{gather}
Plane stress \eqref{plane stress} implies $p=\mu\zeta^{2} $,
and substitution of \eqref{zeta} in the expressions for ${W}$,  ${\boldsymbol{\sigma}}$ and ${\mathbf{P}}$ gives
\begin{gather}
{W}(R) = \frac{\mu}{2}\left(r_{,R}^2(R) + \frac{r^2(R)}{R^2} + \frac{R^2}{r_{,R}^2(R)r^2(R)} -3 \right)\label{SEDF},\\
{\boldsymbol{\sigma}}(R) =\mu\, \mathrm{diag}\left\{
r_{,R}^2(R)- \frac{R^2}{r_{,R}^2(R)r^2(R)}, \frac{r^2(R)}{R^2}-\frac{R^2}{r_{,R}^2(R)r^2(R)}, 0 \right\}, \label{Chauchy stress}
\end{gather}
and
\begin{equation}
{\mathbf{P}}(R) =\mu\, \mathrm{diag}\left\{
r_{,R}(R)- \frac{R^2}{r_{,R}^3(R)r^2(R)}, \frac{r(R)}{R}-\frac{R^3}{r_{,R}^2(R)r^3(R)}, 0 \right\}. \label{Piola stress}
\end{equation}

We note that both the $\theta$ and the $z$ components of the equilibrium equation \eqref{referencial equilibrium} vanish identically, and the radial component is
\begin{equation}
\frac{\partial{P}_{rR}}{\partial R}+ \frac{1}{R}\left({P}_{rR} - {P}_{\theta\Theta}\right)=0.\label{equilimbrium R eq}
\end{equation}
Substituting ${\mathbf{P}}(R)$ into \eqref{equilimbrium R eq}, the non-linear governing equation for the problem is
\begin{equation}
f(r(R),R)\equiv r_{,RR}+\frac{r_{,R}}{R}-\frac{r}{R^{2}}+3\frac{R}{r^{2}r_{,R}^{2}}\left(\frac{Rr_{,RR}}{r_{,R}^{2}}+\frac{R}{r}-\frac{1}{r_{,R}}\right)=0.\label{eq:diff_eq_1}
\end{equation}
The associated boundary conditions \eqref{Boundary Conditions} are
\begin{equation}
r(R_\mathrm{o}) = \lambda_\mathrm{o}R_\mathrm{o},\qquad
\frac{r(R_\mathrm{i})}{R_\mathrm{i}}r_{,R}^2(R_\mathrm{i})=1,
\label{r(R) boundary conditions}
\end{equation}
and where the  conditions on $\sigma_{rz}$ and $\sigma_{\theta z}$ are satisfied identically.
Equations \eqref{eq:diff_eq_1} and \eqref{r(R) boundary conditions} define the boundary value problem (BVP) of the stretched membrane.



In the limit of infinitesimal deformation, $\lambda_\mathrm{o}=1+\epsilon$ where $\epsilon\ll 1$. In this limit the well known solution is \citep[e.g.,][]{Shames1992:1}
\begin{equation}
r = r^\mathrm{S}(R) + O(\epsilon^2) ; \quad r^\mathrm{S}(R) = AR+\frac{B}{R}.\label{small def limit}
\end{equation}
The constants
\begin{equation}
A = 1+\frac{1}{1+3R_\mathrm{i}^2/R_\mathrm{o}^2}\epsilon,\quad
B = \frac{3R_\mathrm{i}^2}{1+3R_\mathrm{i}^2/R_\mathrm{o}^2}\epsilon\label{small def params}
\end{equation}
are obtained from the boundary conditions.

For later reference we define the ratio between the hole and the membrane radii in the deformed configuration as the \emph{hole expansion ratio},
\begin{equation}
\rho \equiv \frac{r_\i}{r_\o}
=\frac{\lambda_\mathrm{i}}{\lambda_\mathrm{o}},
\end{equation}
where $r_\i\equiv r(R_\i)$ and ${\lambda_\mathrm{i}}\equiv\lambda_\theta({R_\mathrm{i}})$ is the tangential stretch at the hole boundary.
In the small deformation limit
\begin{equation}
 \rho = \rho^S \equiv 1+ 3 \frac{R_\o^2-R_\i^2}{R_\o^2+3R_\i^2}\epsilon.
\end{equation}

In order to solve the problem in the limit of \emph{infinite deformation}
it is useful to represent the boundary value problem in 
terms of the dimensionless variables
\begin{equation}
\kappa \equiv \frac{R_\mathrm{i}^2}{R^2}\quad \text{and} \quad \Lambda(\kappa) \equiv \frac{r(R)}{\lambda_\mathrm{o} R} = \frac{\lambda_\theta}{\lambda_\mathrm{o}}. \label{lambda sub}
\end{equation}
Note that the hole expansion ratio is
\begin{equation}
 \rho = \Lambda(1).
\end{equation}

In terms of $\kappa$ and $\Lambda$ the governing equation \eqref{eq:diff_eq_1} takes the simpler form
\begin{equation}\label{lambda eq}
F(\Lambda(\kappa),\kappa,\lambda_\mathrm{o}) = \Lambda_{,\kappa\kappa} + \frac{3}{2\lambda_\mathrm{o}^6}\frac{\left(\Lambda^2\right)_{,\kappa\kappa}}{\Lambda^3(\Lambda-2\kappa \Lambda_{,\kappa})^4}=0.
\end{equation}
The corresponding boundary conditions at the outer boundary and at the hole are
\begin{equation}
\Lambda\left(\kappa_\mathrm{o}\equiv \frac{R_\mathrm{i}^2}{R_\mathrm{o}^2}\right)=1,\quad
\mathrm{and}\quad
\Lambda(1)\left[\Lambda(1)-2\Lambda_{,\kappa}(1)\right]^2=\lambda_\mathrm{o}^{-3},
 \label{labda B.C}
\end{equation}
respectively.

In terms of $\kappa$ and $\Lambda$, the solution at the small deformation limit is
$\Lambda^S(\kappa) = a^S + b^S \kappa$,
where
\begin{equation}
a^S = \frac{A}{\lambda_\mathrm{o}}=1-\frac{3\kappa_\mathrm{o}}{1+3\kappa_\mathrm{o}}\epsilon,
\quad
\text{and}\quad
b^S  = \frac{B}{\lambda_\mathrm{o} R_\mathrm{i}^2}=\frac{3}{1+3\kappa_\mathrm{o}}\epsilon.\label{small deformation a,b}
\end{equation}
Adopting a similar form we define
\begin{equation}\label{lambda small def}
 \Lambda_\mathrm{L}(\kappa) = a + b \kappa,
\end{equation}
and substitute for $\Lambda$ into equation \eqref{lambda eq}, the resulting term is
\begin{equation}\label{equation subs}
F_\mathrm{L}(\Lambda_\mathrm{L}(\kappa),\kappa,\lambda_\mathrm{o}) 
= \frac{3 b^2}{\lambda_\mathrm{o}^6 (a-b \kappa)^4 (a+b \kappa)^3}.
\end{equation}
Since $F_\mathrm{L}\neq 0$, \eqref{lambda small def} is generally not a solution to the problem.
However, as $\lambda_\mathrm{o}\rightarrow \infty$ we have that $F_\mathrm{L}\to 0$, suggesting that \eqref{lambda small def} is the solution to the problem in the infinite deformation limit.
The boundary condition on the outer radius leads to
\begin{equation}
\label{aofb}
 a = 1 - b\kappa_\o,
\end{equation}
and at the boundary of the hole, in the limit $\lambda_\o \to \infty$, we have that
\begin{equation}
 \lim \limits_{\lambda_\o \to \infty} b = b_\infty \equiv \frac{1}{1+\kappa_\o},
\end{equation}
and $a_\infty$ is obtained by substituting $b_\infty$ in \eqref{aofb}.
This leads to the following expression for $\Lambda_\mathrm{L}$ in the limit $\lambda_\o \to \infty$, namely
\begin{equation} \label{inf y sol}
\Lambda^\infty(\kappa)=\frac{1+\kappa}{1+\kappa_\mathrm{o}}.
\end{equation}

Before we proceed, we note that by substituting $\Lambda^\infty$ back in \eqref{equation subs} we have that
\begin{equation}
F(\Lambda^\infty(\kappa),\kappa,\lambda_\mathrm{o}) 
= \frac{3 (1+\kappa_\mathrm{o})^5}{\lambda_\mathrm{o}^6 (1-\kappa)^4 (1+\kappa)^3}.\label{f in infinite deformation}
\end{equation}
In the limit $\lambda_\mathrm{o}\rightarrow \infty$, this expression vanishes for any $\kappa \neq 1$, that is everywhere but at the inner boundary.
Thus, $\Lambda^\infty$ is the solution for the problem in the infinite deformation limit, with an identified singularity at the inner boundary.
We further note that if we consider a series solution for the problem in the form
\begin{equation}
 \Lambda = a_\infty + b_\infty \kappa + b_p(\kappa) \left(\frac{1}{\lambda_\o}\right)^{p},
\end{equation}
as $\lambda_\o\rightarrow\infty$ the lowest non vanishing correction term is $p = 3/2$.
This higher than linear correction term implies that $\Lambda^\infty$ approximates the exact solution for an extremely wide range of deformations.

Returning to physical variables, the solution for $r(R)$ in the infinite deformation limit is
\begin{equation}
 r^{\infty}(R) = \lambda_\o R_\o \frac{\sfrac{R}{R_\i}+\sfrac{R_\i}{R}}{\sfrac{R_\o}{R_\i}+\sfrac{R_\i}{R_\o}}.
\label{radius infinite stretch}
\end{equation}
%
The hole expansion ratio in this limit is
\begin{equation}\label{strech ratio limit}
\rho^{\infty} = 
\Lambda^\infty(1) 
=\frac{2}{1+\kappa_\mathrm{o}} = \frac{2 R_\o^2}{R_\o^2 + R_\i^2}.
\end{equation}

Next, we propose an approximation for a range of deformations which are finite but rather moderate.
This is accomplished by adding a quadratic term in $\kappa$ to the small deformation solution.
This leads to an expression in the form a Taylor expansion series.
\begin{equation}
\tilde{\Lambda}^\mathrm{S}(\kappa) = a^\mathrm{S}+b^\mathrm{S}\kappa+\gamma^\mathrm{S} \kappa^2 = \tilde{\rho}^\mathrm{S} + \beta^\mathrm{S}(\kappa-1)+\gamma^\mathrm{S}(\kappa-1)^2, \label{quadratic  correction}
\end{equation}
where we find the second expression in \eqref{quadratic  correction} more convenient for the subsequent analysis,
where $\tilde{\rho}^\mathrm{S} = a^\mathrm{S}-\beta^\mathrm{S}+\gamma^\mathrm{S}$
and $\beta^\mathrm{S} = b ^\mathrm{S} + 2\gamma^\mathrm{S}$.
In order for $\tilde{\Lambda}^\mathrm{S}$ to approach $\Lambda^\mathrm{S}$ in the small deformation limit we require that
\begin{equation}
\lim \limits_{\lambda_\mathrm{o} \rightarrow 1}\gamma^\mathrm{S} = 0.\label{gamma constrint}
\end{equation}
Note that $\tilde{\Lambda}^\mathrm{S}(1)=\tilde{\rho}^\mathrm{S}$ is an approximation for the hole expansion ratio.
Therefore, we have the additional constraint
\begin{equation}
\lim \limits_{\lambda_\mathrm{o} \rightarrow 1}\tilde{\rho}^\mathrm{S} = 1.\label{rho constraint}
\end{equation}

The approximation we propose is constructed such that it will satisfy the differential equation \eqref{lambda eq} at the inner boundary ($\kappa=1$), and we assume that $F$ will remain small away from it.
Thus,
\begin{gather}
F(\tilde{\Lambda}^\mathrm{S}(1),1,\lambda_\mathrm{o}) =\frac{(3 \beta^\mathrm{S}) ^2}{\lambda_\mathrm{o}^6 (\tilde{\rho}^\mathrm{S})^3 (\tilde{\rho}^\mathrm{S}-2 \beta^\mathrm{S} )^4}
+\gamma^\mathrm{S}  \left(\frac{6}{\lambda_\mathrm{o}^6 (\tilde{\rho}^\mathrm{S})^2 (\tilde{\rho}^\mathrm{S}-2 \beta^\mathrm{S} )^4}+2\right)=0
\end{gather}
leads to
\begin{gather}
\gamma^\mathrm{S} = -\frac{3 (\beta^\mathrm{S}) ^2}{2 \tilde{\rho}^\mathrm{S}  \left[\lambda_\mathrm{o}^6 (\tilde{\rho}^\mathrm{S}) ^2 (\tilde{\rho}^\mathrm{S} -2 \beta^\mathrm{S} )^4+3\right]}.
\end{gather}
The boundary condition at the hole requires
\begin{equation}\label{beta S}
\beta^\mathrm{S} = \frac{\tilde{\rho}^\mathrm{S}}{2}\pm\frac{1}{2\sqrt{\tilde{\rho}^\mathrm{S}}\lambda_\mathrm{o}^{3/2}}.
\end{equation}
Since only the $(-)$ option in \eqref{beta S} satisfies \eqref{gamma constrint}, we have that
\begin{equation}
\beta^\mathrm{S} = \frac{\tilde{\rho}^\mathrm{S}}{2}-\frac{1}{2\sqrt{\tilde{\rho}^\mathrm{S}}\lambda_\mathrm{o}^{3/2}}, \quad
\gamma^\mathrm{S} =- \frac{3\left[(\tilde{\rho}^\mathrm{S})^{3/2}\lambda_\mathrm{o}^{3/2}- 1\right]^2}{32(\tilde{\rho}^\mathrm{S})^2\lambda_\mathrm{o}^3}. \label{beta gamma constraints}
\end{equation}
In order to satisfy the boundary condition at the outer radius, $\lambda_\o$ must be related to $\tilde{\rho}^\mathrm{S}$ by
\begin{equation}
 \lambda_\o = \mathcal{L}_{\kappa_\o}(\tilde{\rho}^S),
\end{equation}
where
\begin{equation}
 \mathcal{L}_{\kappa_\o}(\rho) = \left(\frac{1}{\rho^2}\frac{3^2(1-\kappa_\o)^2}
    {
    (281-66 \kappa_\o + 9 \kappa_\o^2)\rho
     - 8 \sqrt{2}(3\kappa_\o - 11)\sqrt{\rho(5\rho-1)}
     - 96
    }\right)^{\frac{1}{3}}.
\end{equation}
Since only one branch of $\mathcal{L}_{\kappa_\o}(\rho)$ satisfies \eqref{rho constraint}, $\tilde{\rho}^\mathrm{S}$ can be determined uniquely by the inverse of the said branch,
\begin{equation}
 \tilde{\rho}^S = \mathcal{L}_{\kappa_\o}^{-1} (\lambda_\o)
\end{equation}
We note that since $\lim \limits_{\lambda_\o \to \infty} \mathcal{L}_{\kappa_\o}^{-1}(\lambda_\o)>\rho^\infty$,
at large deformations this approximate solution does not converge to the solution of this problem.

When the punctured membrane is subjected to large deformation we distinguish between the responses in the inner and outer regions.
Specifically, we assume that in the vicinity of the hole there is a band in which the dependence of the tangential stretch on $R$ is small in comparison with the corresponding dependence in the outer part.
We further assume that this band shrinks as the applied stretch increases and finally, in the limit $\lambda_\o \to \infty$, reduces to the singularity at the surface of the hole.
Note that this model resembles the partition of a wrinkled membrane to an inner wrinkly and an outer taut regions \cite{Wu1978}.

According to our assumption $r^\infty$ provides a good approximation for the solution in the outer region.
Indeed near the outer boundary, in the region $R_\mathrm{o}\geq R\gg R_\mathrm{i}$,
\begin{equation}
f(r^\infty(R),R)=\frac{1}{\lambda_\mathrm{o}^6}\left(12 \left(\frac{R_\mathrm{i}}{R}\right)^4 + O\left(\frac{R_\mathrm{i}}{R}\right)^6\right) \ll 1
\end{equation}
is small as long as $\lambda_\o$ is large.
Therefore, $r^\infty$ can serve as a basis for an approximate solution near the outer boundary in the large deformation regime.

Following this observation, our approximation in the outer region will be based on the infinite deformation solution \eqref{inf y sol} with a quadratic correction term in $\kappa$.
In the inner region we approximate the solution with a quadratic polynomial in the form of \eqref{quadratic  correction}.
At the interface between the inner and outer section we require continuity of the radial deformation and stress.
These are fulfilled by requiring continuity of $r(R)$ and its first derivative.

Specifically, in terms of $\kappa$ and $\Lambda$,
the proposed large deformations approximation is
\begin{equation}
\tilde{\Lambda}^{\infty}
=
 \begin{cases}
\tilde{\Lambda}^\mathrm{I} = \tilde{\rho}^\infty +\beta^\infty(\kappa-1)+\gamma^\infty (\kappa-1)^2,	& 1\geq \kappa \geq \kappa_\mathrm{t}\\
\tilde{\Lambda}^\mathrm{O} = \frac{1+\kappa}{1+\kappa_\mathrm{o}}+c(\kappa-\kappa_\mathrm{o})^2,	&\kappa_\mathrm{t}\geq \kappa\geq \kappa_\mathrm{o}
\end{cases},
\end{equation}
where $\kappa_\mathrm{t} \equiv (R_\i/R_\mathrm{t})^2$ and $R_\mathrm{t}$ is the transition radius between the inner and outer sections.

The proposed approximation involve six unknowns, the five constants and the transition radius $R_\mathrm{t}$.
In order to determine these unknowns,
two equations are obtained by imposing the governing equation to vanish at the boundary of the hole and the outer boundary.
Additional two equations arise from the boundary conditions.
The final two equations are obtained from the continuity requirement on the deformation and its derivative.

Note that $\tilde{\Lambda}^\mathrm{I}$ is subjected to the same conditions at the inner boundary as $\tilde{\Lambda}^\mathrm{S}$.
Hence, the relations between $\beta^\mathrm{S},\gamma^\mathrm{S}$ and $\tilde{\rho}^\mathrm{S}$ in \eqref{beta gamma constraints} apply for $\beta^\infty,\gamma^\infty$ and $\tilde{\rho}^\infty$ too.
Similarly to $\tilde{\rho}^\mathrm{S}$, $\tilde{\Lambda}^\infty(1)=\tilde{\rho}^\infty$ is an approximation for the hole expansion ratio.

In the outer region the condition we impose on $\tilde{\Lambda}^\mathrm{O}$ is that $F$ vanishes at the outer boundary $(\kappa=\kappa_\mathrm{o})$, assuming that it will remain small away from it.
Thus,
\begin{gather}
F\left(\tilde{\Lambda}^{\mathrm{O}}(\kappa_\o),\kappa_\mathrm{o},\lambda_\mathrm{o}\right) =
2 c+ \frac{3 (\kappa_\mathrm{o}+1)^2 \left(2 c (\kappa_\mathrm{o}+1)^2+1\right)}{\lambda_\mathrm{o}^6 (\kappa_\mathrm{o}-1)^4}=0
\end{gather}
leads to
\begin{gather}
c = -\frac{3 (\kappa_\mathrm{o}+1)^2}{2 \lambda_\mathrm{o}^6 (\kappa_\mathrm{o}-1)^4+6 (\kappa_\mathrm{o}+1)^4}.
\end{gather}

The requirement of the smooth transition between the two parts of the approximate solution allows to obtain $\kappa_\mathrm{t}$ as a function of $\lambda_\mathrm{o}$ and $\tilde{\rho}^\infty$.
Specifically,
\begin{gather}
\kappa_\mathrm{t} =\frac{1}{c-\gamma^\infty}\left( \frac{\beta(1+\kappa_\mathrm{o})-1}{1+\kappa_\mathrm{o}}+c\kappa_\mathrm{o}-\gamma^\infty\right). \label{smoothness, XI}
\end{gather}
Finally, continuity of displacement dictates the relation between $\tilde{\rho}^\infty$ and $\lambda_\mathrm{o}$ via the implicit relation
\begin{equation}
\tilde{\rho}^\infty + \beta^\infty(\kappa_\mathrm{t}-1)+\gamma^\infty(\kappa_\mathrm{t}-1)^2 = \frac{1+\kappa_\mathrm{t}}{1+\kappa_\mathrm{o}}+c(\kappa_\mathrm{t}-\kappa_\mathrm{o})^2.
\label{continuity}
\end{equation}
Note that not all values of $\lambda_\mathrm{o}\geq 1$ can be matched with a $1 \leq \tilde{\rho}^\infty<2/(1+\kappa_\mathrm{o})$ such that \eqref{continuity} is satisfied.
Moreover, not all pairs of $\lambda_\mathrm{o},\tilde{\rho}^\infty$ that satisfy \eqref{continuity} correspond to a value of $\kappa_\mathrm{o}\leq \kappa_\mathrm{t}\leq 1$.
Next, we determine the range of $\lambda_\o$ for which the large deformation approximation is valid.

If the applied stretch is gradually increased from $\lambda_\o=1$ there is a \emph{transition stretch} $\lambda_\mathrm{o}^{\mathrm{\tau}}$ at which the membrane response switches from intermediate to large.
Moreover, when $\lambda_\o=\lambda_\mathrm{o}^{\mathrm{\tau}}$ then
$\kappa_\mathrm{t}=\kappa_\mathrm{o}$ and hence
$\tilde{\Lambda}^\mathrm{S}=\tilde{\Lambda}^\mathrm{I}$
as both sencod order polynomials satisfy the same conditions.
The corresponding hole expansion ratio at $\lambda_\o^\tau$ is then
$\tilde{\rho}^\mathrm{\tau}=\tilde{\rho}^{S}(\lambda_\mathrm{o}=\lambda_\mathrm{o}^\tau) = \tilde{\rho}^{\infty}(\lambda_\mathrm{o}=\lambda_\mathrm{o}^\tau)$.
The transition parameters $\lambda_\mathrm{o}^\mathrm{\tau}$ and $\tilde{\rho}^\mathrm{\tau}$ are found by solving the continuity \eqref{continuity} and smoothness \eqref{smoothness, XI} conditions on $\tilde{\Lambda}^\infty$ in the limit
$\kappa_\mathrm{t} \to \kappa_\mathrm{o}$.
In this limit, \eqref{smoothness, XI} becomes
\begin{gather}
\frac{2 (3 \kappa_\mathrm{o}-7)}{\lambda_\mathrm{o}^{3/2} \sqrt{\tilde{\rho}^\infty }}+\frac{3-3 \kappa_\mathrm{o}}{\lambda_\mathrm{o}^3 (\tilde{\rho}^\infty) ^2}+\tilde{\rho}^\infty  (11-3 \kappa_\mathrm{o})-\frac{16}{\kappa_\mathrm{o}+1}=0.
\end{gather}
This quadratic equation has two solutions for $\lambda_\mathrm{o}^\mathrm{\tau}$,
out of which only
\begin{equation}
\lambda_\mathrm{o}^\mathrm{\tau}=\frac{3^{2/3} (1-\kappa_\mathrm{o})^{2/3}}{\tilde{\rho}^\mathrm{\tau}} \left(7-3 \kappa_\mathrm{o} - 4 \sqrt{1+3\frac{1-\kappa_\mathrm{o}}{\tilde{\rho}^\mathrm{\tau}(1+\kappa_\mathrm{o})}}\right)^{-\frac{2}{3}}, \label{labda_o^t 1}
\end{equation}
corresponds to $\lambda_\mathrm{o}^\mathrm{\tau}>1$.
Substituting $\lambda_\mathrm{o}^\mathrm{\tau}$ in \eqref{continuity} and taking the limit $\kappa_\mathrm{t}\rightarrow \kappa_\mathrm{o}$ gives
\begin{equation}
\kappa_\mathrm{o} \left[2 \tilde{\rho}^\mathrm{\tau}  \left(\sqrt{1+3\frac{1-\kappa_\mathrm{o}}{\tilde{\rho}^\mathrm{\tau}(1+\kappa_\mathrm{o})}}-4\right)+3\right]+
2 \tilde{\rho}^\mathrm{\tau}  \left(\sqrt{1+3\frac{1-\kappa_\mathrm{o}}{\tilde{\rho}^\mathrm{\tau}(1+\kappa_\mathrm{o})}}-4\right)+9=0.
\end{equation}
From the two solutions for $\tilde{\rho}^\mathrm{\tau}$, only
\begin{equation}
\tilde{\rho}^\mathrm{\tau} = \frac{3 \kappa_\mathrm{o}+13 +\sqrt{34-6 \kappa_\mathrm{o} (\kappa_\mathrm{o}+2)}}{10 (\kappa_\mathrm{o}+1)},
\end{equation}
that corresponds to $1<\tilde{\rho}^\mathrm{\tau}<2/(1+\kappa_\mathrm{o})$ is feasible.
Substituting $\tilde{\rho}^\mathrm{\tau}$ back to \eqref{labda_o^t 1} results in
\begin{equation}
\begin{aligned}
\lambda_\mathrm{o}^\mathrm{\tau} &= \frac{10\cdot 3^{2/3} (\kappa_\mathrm{o}+1)(1-\kappa_\mathrm{o})^{2/3}}{3 \kappa_\mathrm{o}+\sqrt{34-6 \kappa_\mathrm{o} (\kappa_\mathrm{o}+2)}+13}\\
&~~
\left(7-3 \kappa_\mathrm{o}-4 \sqrt{\frac{43 -27 \kappa_\mathrm{o}+\sqrt{34-6 \kappa_\mathrm{o} (\kappa_\mathrm{o}+2)}}{13+3 \kappa_\mathrm{o}+\sqrt{34-6 \kappa_\mathrm{o} (\kappa_\mathrm{o}+2)}}}\right)^{-2/3}.
\end{aligned}
\label{transition stretch}
\end{equation}
Thus, we obtained a closed-form expression for the transition stretch from the intermediate to the large deformation regime.
This depends solely on the geometry of the membrane in terms on the ratio between the hole and the membrane radii $\kappa_\o = (R_\i/R_\o)^2$.


Of particular interest is the asymptotic limit of an infinitesimally small hole.
In the limit of infinite deformation, according to Eq.~\eqref{strech ratio limit} the hole expansion ratio is $\rho^{\infty}=2$. 
That is, under plane stress the radius of the hole is twice that of a similar hole in a membrane under the aforementioned plane strain condition.
Yet, Eq.~\eqref{radius infinite stretch} implies that away from the hole, where $R\gg R_\i$, the solution rapidly converges to the plane-strain case.

For small holes the transition stretch from intermediate to large deformations 
\begin{equation}
 \lambda_\o^\tau \to 1.628 + 1.954 \kappa_\o.
\end{equation}
is obtained via a Taylor series expansion of \eqref{transition stretch}.
At this transition stretch the hole expansion ratio is
\begin{equation}
 \tilde{\rho}^{\tau} \to 1.883 - 1.686 \kappa_\o.
\end{equation}
We note that the hole expansion ratio varies only slightly as the membrane stretch increases from $\lambda_\o^\tau$ to infinity (form 1.883 to 2).

\begin{figure}[t]
\centering
\includegraphics[scale=0.73]{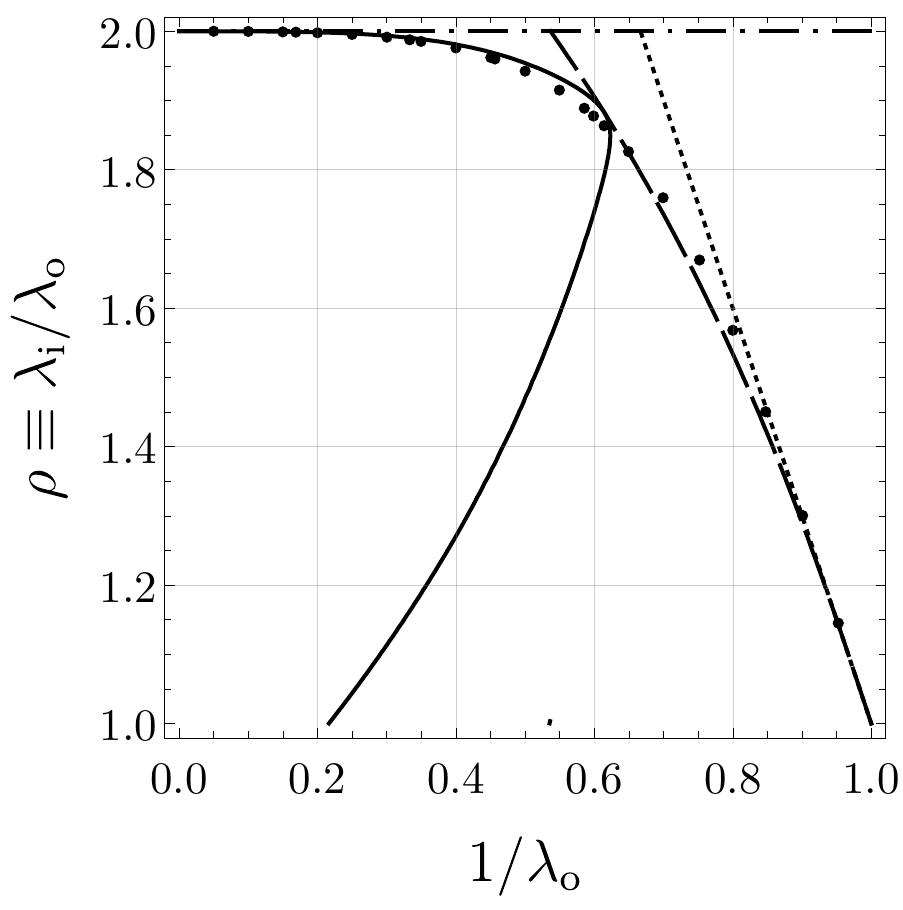}
\caption{\small{The hole expansion ratio as a function of $1/\lambda_\o$ in the limit of infinitesimally small hole according to
the numerical results (dots), the infinite and small deformation limits (dash-dotted and densely dashed lines, respectively), and the intermediate and large deformation approximations, (dashed and continuous curves, respectively).}}\label{lambdao,rho X_o_0}
\end{figure}
Fig.~\ref{lambdao,rho X_o_0} shows the variation of the hole expansion ratio
$\rho$ as a function of $1/\lambda_\mathrm{o}$, according to the intermediate and large deformation approximations (dashed and continuous curves, respectively), together with extensions of the solutions for the small and infinite deformation limits (densely dashed and continuous lines, respectively).
Also shown are the corresponding results determined by numerical solution of the governing Eq.~\eqref{lambda eq} (dots). 
Interestingly, we note that the form of the governing equation \eqref{lambda eq} enables to determine a numerical solution for the problem even in the limit $R_\i\rightarrow 0$ where $0\leq\kappa\leq 1$.

We observe that the intermediate and large approximations agree at a single point, ($\lambda_\o = 1.628$ and $\tilde{\rho} = 1.883$).
Moreover, the joint curve of the two approximations in their respective domains of applicability neatly agrees with the corresponding numerical results.
Thus, usage of $\tilde{\rho}^\mathrm{S}$ for $\lambda_\mathrm{o}<\lambda_\mathrm{o}^\mathrm{\tau}$ and $\tilde{\rho}^\infty$ for $\lambda_\mathrm{o}>\lambda_\mathrm{o}^\mathrm{\tau}$ 
provides a neat approximation for the numerical results for $\rho$ throughout the entire range of deformations.

\section{Numerical Application}
We begin this section with a through comparison between the proposed approximation and the corresponding numerical solution of the problem.
For conciseness, in the sequel we make use of the intermediate and the large approximations in their perspective ranges of the applied stretch.

\begin{figure}[t]
\centering
\includegraphics[scale=0.8]{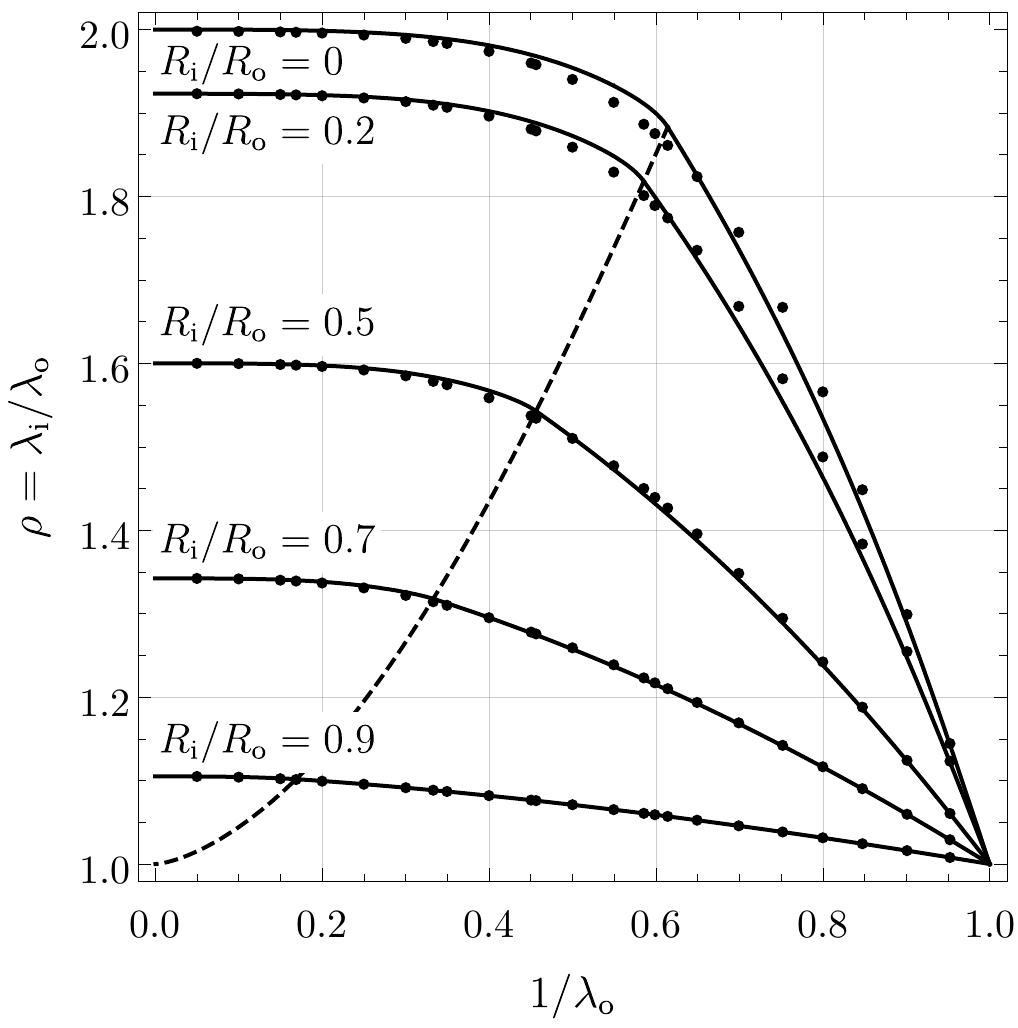}
\caption{\small{The hole expansion ratio as a function of $1/\lambda_\mathrm{o}$ for several values of $\sqrt{\kappa_\mathrm{o}} = R_\mathrm{i}/R_\mathrm{o}$.
The continuous curves and dots correspond to the approximation and the numerical results, respectively.
The dashed curve depicts the transition stretch $\lambda_\o^\tau$ as a function of $R_\i/R_\o$}}\label{rho(1/lambda) curves}
\end{figure}
Fig.~\ref{rho(1/lambda) curves} shows the variation of the hole expansion ratio as a function of $1/\lambda_\mathrm{o}$ for several values of $R_\i/R_\o$, according to the proposed approximation (continuous curves) and numerical results (dots).
The dashed curve is a parametric curve that marks the transition between the intermediate deformation approximation $\tilde{\Lambda}^S$ and the large deformation approximation $\tilde{\Lambda}^\infty$.
We observe the  agreement between the analytical approximation and the numerical results throughout the entire range of stretch ratios and hole sizes.

\begin{figure}[t]
\centering
 \includegraphics[width=0.43\linewidth]{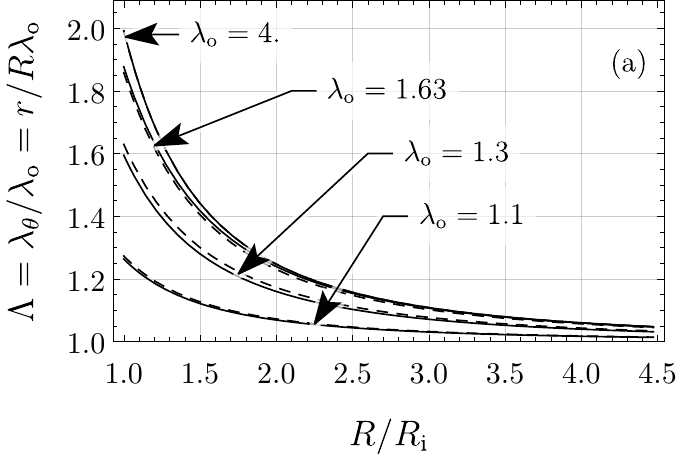}
 \quad
 \includegraphics[width=0.43\linewidth]{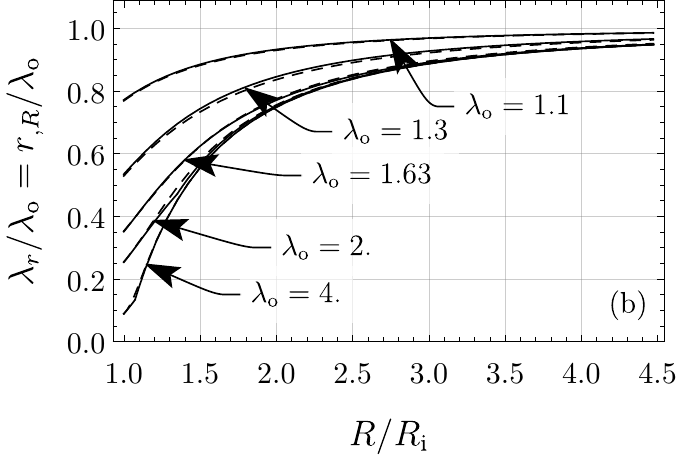}
\caption{\small{Principal stretches in the punctured membrane for $R_\mathrm{i} = R_\mathrm{o}/20$.
Continuous and dashed curves correspond to approximate and numerical results, respectively.}}\label{comparison}
\end{figure}
Fig.~\ref{comparison} compares between the proposed approximation and the numerical results for 
the tangential and radial stretches in the membrane in the case
$\sqrt{\kappa_\mathrm{o}}=R_\mathrm{i}/R_\mathrm{o} = 1/20$ and various values of $\lambda_\mathrm{o}$.
The largest difference between the approximation and the numerical solution, which is smaller than $5\%$, is exhibited at $\lambda_\o = 1.3$.
Both Figs.~\ref{rho(1/lambda) curves} and \ref{comparison} demonstrate that the approximation proposed herein agrees with the numerical solutions throughout the entire range of deformations and ratios between the hole and the membrane radii.

\begin{figure}[b]
 \centering
 \begin{subfigure}[b]{0.48\textwidth}
  \includegraphics[height=0.6\linewidth]{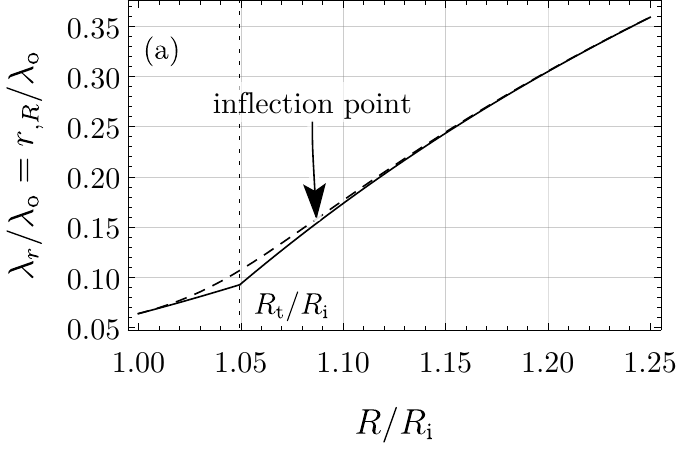}
 \end{subfigure}
 \begin{subfigure}[b]{0.48\textwidth}
  \includegraphics[height=0.63\linewidth]{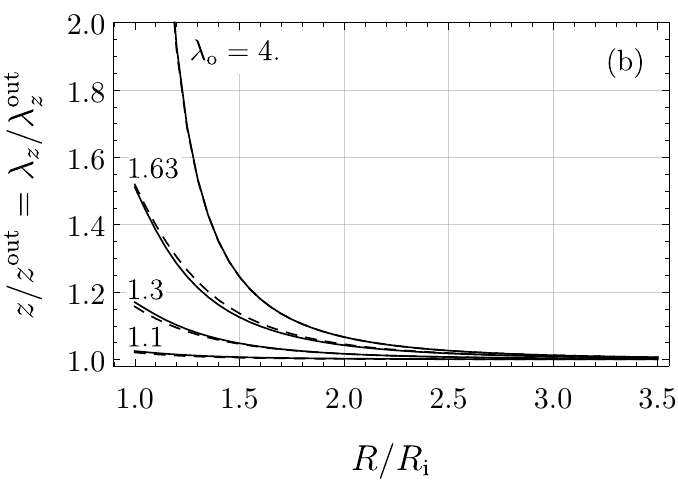}
 \end{subfigure}
 \caption{\small{Normalized radial (a) and axial (b) stretches in the vicinity of a small hole of a punctured membrane ($R_\i = R_\o/20$). The continuous and dashed curves correspond to the approximation and the numerical results, respectively. The curve in (a) corresponds to $\lambda_\o = 5$.}}\label{near hole}
\end{figure}

The distribution of radial stretch near a small hole ($R_\i = R_\o/20$) is shown for $\lambda_\o = 5$ in Fig.~\ref{near hole}a.
This serves to demonstrate the change in the trend of the solution in the vicinity of the hole at large stretches.
In the numerical results we observe an inflection point at which the variation rate of the radial stretch is maximal.
The lower slope of the curve near the hole is captured by $\tilde{\Lambda}^I$.
We further note that both the inflection point and the transition radius $R_\mathrm{t}$ approach the hole boundary as the applied stretch increases.

In the limit of an infinitesimally small hole under infinite deformation, the axial stretch at the surface of the hole is $\lambda_z = 1/\sqrt{2\lambda_\o}$, and the axial stretch at the membrane outer boundary is $\lambda_z^\mathrm{out} = 1/\lambda_\o^2$. 
Clearly, the membrane gets thinner everywhere. Yet, the thickness at the surface of the hole becomes unboundedly larger relative to the thickness at away from the hole.
Fig.~\ref{near hole}b illustrates this phenomenon for a small hole for which $R_\i = R_\o/20$.
As was mentioned before, we note that at radii larger than $3R_\i$, the membrane becomes virtually flat, implying that plane strain condition is applicable away from the hole.
Nonetheless, the rapid variation in the thickness near the hole is fundamentally different from the uniform deformation determined by
\cite{haughton1991exact} for the punctured Verga membrane.


\begin{figure}[t]
\centering
\includegraphics[scale=0.9]{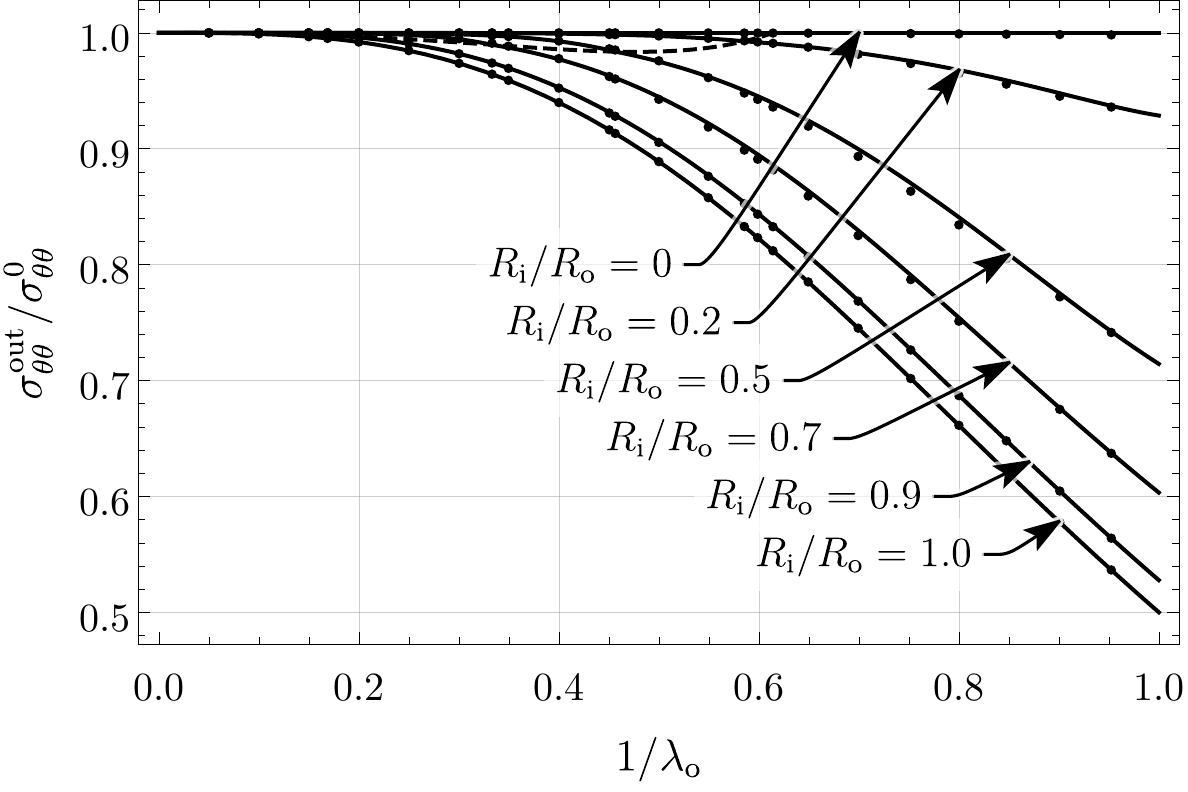}
\caption{\small{The hoop stress at the outer boundary normalized by the nominal stress as a function of $1/\lambda_\mathrm{o}$ for several values of $R_\i/R_\o$.
The continuous curves and dots correspond to the approximation  and the numerical results, respectively.
The dashed curve depicts the transition stretch $\lambda_\o^\tau$ as a function of $R_\i/R_\o$.}}\label{sigma theta theta}
\end{figure}
Fig.~\ref{sigma theta theta} shows the ratio between the hoop stress at the outer boundary and the stress in a membrane with no hole (the nominal stress) as a function of $1/\lambda_\mathrm{o}$ for different values of $\kappa_\mathrm{o}$. 
Here and elsewhere, we compare the analytical results in the limit $R_\i/R_\o\rightarrow1$ with the numerical results for $R_\i/R_\o=0.99$.
We note that for any hole size, the difference between $\sigma_{\theta \theta}^0$ and $\sigma_{\theta \theta}^\mathrm{out}$ becomes negligible for $\lambda_\mathrm{o}>\lambda_\mathrm{o}^\tau$.
Furthermore for small holes, regardless of the applied stretch, the hoop stress at the outer boundary is practically identical to the stress in a membrane with no hole, that is
\begin{equation}
\lim \limits_{\kappa_\mathrm{o}\rightarrow 0} \sigma_{\theta \theta}^\mathrm{out}= \sigma_{\theta \theta}^0 = \mu \left(\lambda_\mathrm{o}^2 - \lambda_\mathrm{o}^{-4}\right).
\end{equation}
This is not the case for finite holes
in the small deformation limit, where the hoop stress at the outer boundary 
depends on the hole size via the relation
\begin{equation}
\sigma_{\theta \theta}^\mathrm{out,S} = \frac{1+\kappa_\mathrm{o}}{1+3\kappa_\mathrm{o}}6\mu \epsilon.
\end{equation}
However, in the infinite deformation limit,
\begin{equation}
\sigma_{\theta \theta}^{\mathrm{out},\infty} = \mu \lambda_\mathrm{o}^2,
\end{equation}
which is equal $\sigma_{\theta \theta}^0$ in this limit independently of $\kappa_\mathrm{o}$.
In the limit of a thin strip $(R_\i/R_\o \rightarrow 1)$, the hoop stress at both the outer boundary and the hole is
\begin{equation}
 \sigma_u = \mu \left(\lambda_\o^2 - \frac{1}{\lambda_\o}\right),
\end{equation}
which is nothing but the longitudinal stress in a uniaxially stretched strip.

\begin{figure}[t]
\centering
\includegraphics[scale=0.9]{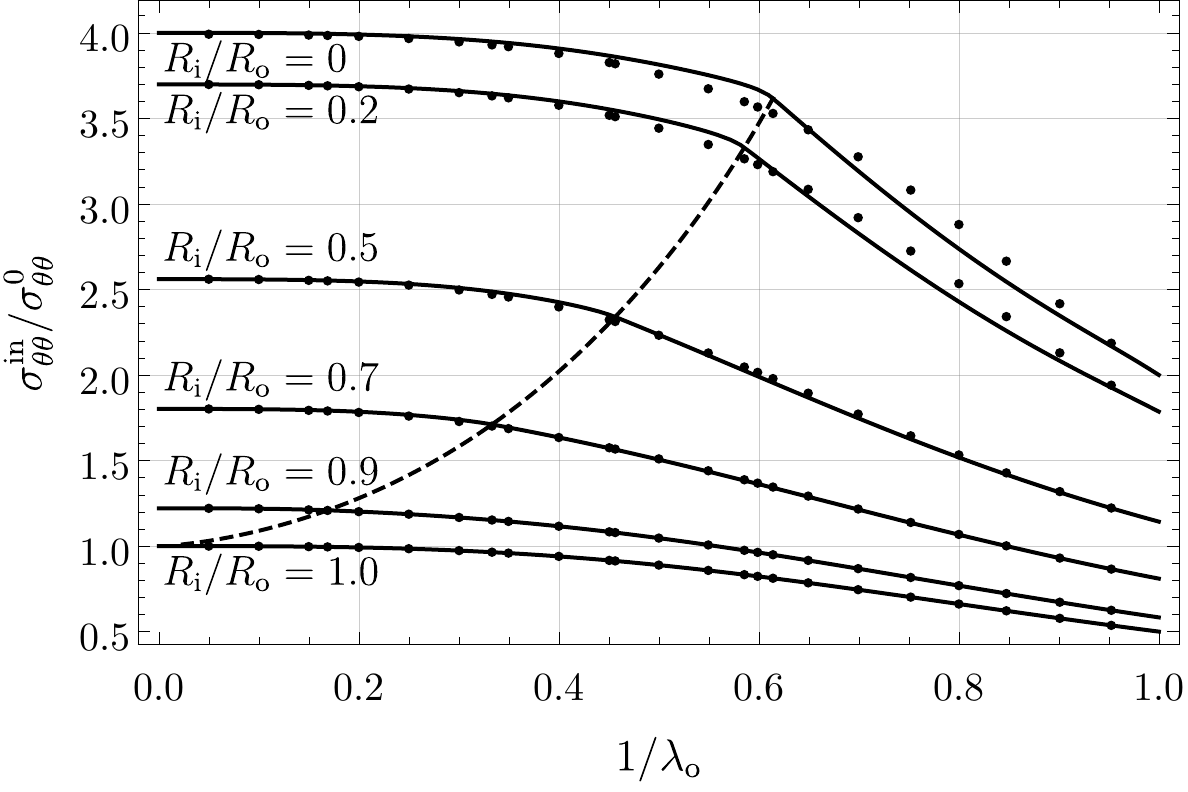}
\caption{\small{The hoop stress at the surface of the hole normalized by the nominal stress as a function of $1/\lambda_\mathrm{o}$ for several values of $R_\i/R_\o$.
The continuous curves and dots correspond to the approximation  and the numerical results, respectively.
The dashed curve depicts the transition stretch $\lambda_\o^\tau$ as a function of $R_\i/R_\o$.}}\label{sigma theta theta ratio}
\end{figure}
Fig.~\ref{sigma theta theta ratio} shows the ratio between
the hoop stress at the inner boundary and the nominal stress as a function of $1/\lambda_\mathrm{o}$ for different values of $\kappa_\mathrm{o}$.
For values of $\lambda_\mathrm{o}$ slightly larger than $\lambda_\mathrm{o}^\tau$ the difference between this ratio and its limit at infinite deformation becomes negligible.
In the limit of a small hole, $\sigma_{\theta \theta}^\mathrm{in}/\sigma_{\theta \theta}^\mathrm{0}$ is known to be 2 in small strains.
In the limit of infinite deformation the hoop stress at the surface of the hole is
\begin{equation}
\sigma_{\theta \theta}^{\mathrm{in},\infty} = \left(\frac{2}{1+\kappa_\mathrm{o}}\right)^2 \mu \lambda_\mathrm{o}^2.
\end{equation}
Comparing to the nominal stress, the stress concentration factor at the surface of the hole in this limit is
\begin{equation}
 \lim \limits_{\lambda_\o \to \infty} \sigma_{\theta \theta}^{\mathrm{in}}/\sigma_{\theta \theta}^{0} = \left(\frac{2}{1+\kappa_\mathrm{o}}\right)^2.
\end{equation}
In the limit of a small hole this ratio approaches 4, twice the stress concentration factor in the limit of infinitesimal deformations.

We consider next the strain energy stored in the punctured membrane.
In the limit of small deformations, the dependence of the energy density function on the radius is
\begin{equation}
{W}^\mathrm{S} = \mu \frac{6\epsilon^2}{\left(1+ 3\kappa_\mathrm{o} \right)^2}\left(1+3\frac{R_\mathrm{i}^4}{R^4}\right) 
.
\end{equation}
Therefore the total strain energy in this limit is
\begin{equation}
{\mathcal{W}}^\mathrm{S} =\int_V {W}^\mathrm{S}\ \mathrm{d}V  = V_0 (1-\kappa_\o) \frac{6\mu\epsilon^2}{1+3\kappa_\mathrm{o}} 
,
\end{equation}
where $V_0 = \pi H R_\o^2$ is the volume of a membrane without a hole and the same outer radius and thickness.
The referential volume of the hole is
\begin{equation}
 V_\mathrm{hole} \equiv V_0 - V = \kappa_\o V_0,
\end{equation}
where $V$ is the volume of the punctured membrane.

The energy density of a membrane without a hole under similar boundary conditions is
\begin{equation}
{W}_0^\mathrm{S} = 6\mu \epsilon^2 
,
\end{equation}
and the total energy stored in the whole membrane is
\begin{equation}
{\mathcal{W}}_0^\mathrm{S} = \int_{V_0} {W}_0^\mathrm{S}\ \mathrm{d}V= 6\mu V_0 \epsilon^2 
.
\end{equation}
The difference between the energies is
\begin{equation}
\Delta_W^S \equiv
{\mathcal{W}}_0^\mathrm{S}-{\mathcal{W}^\mathrm{S}} = \frac{24\kappa_\o}{1+3\kappa_\o}\mu V_0 \epsilon^2 
 = \frac{24}{1+3\kappa_\o}\mu V_\mathrm{hole} \epsilon^2 
.
\end{equation}
In the limit of a small hole this difference becomes
\begin{equation}
\lim_{\kappa_\o \to 0} \Delta_W^S = 24\mu V_\mathrm{hole}\epsilon^2
=4V_\mathrm{hole} W_0^\mathrm{S}
.
\end{equation}

Following similar steps, in the infinite deformation limit the energy difference is
\begin{equation}
\Delta_W^\infty \equiv {\mathcal{W}}_0^\infty-{\mathcal{W}^\infty} = \frac{2}{1+\kappa_\mathrm{o}}\mu V_\mathrm{hole}\lambda_\mathrm{o}^2 
,
\end{equation}
and in the limit of a small hole,
\begin{equation}
\lim_{\kappa_\o \to 0} \Delta_W^\infty =
2\mu V_\mathrm{hole}\lambda_\mathrm{o}^2
=2V_\mathrm{hole}W_0^\infty
.
\end{equation}

\begin{figure}[t]
\centering
\includegraphics[scale = 0.8]{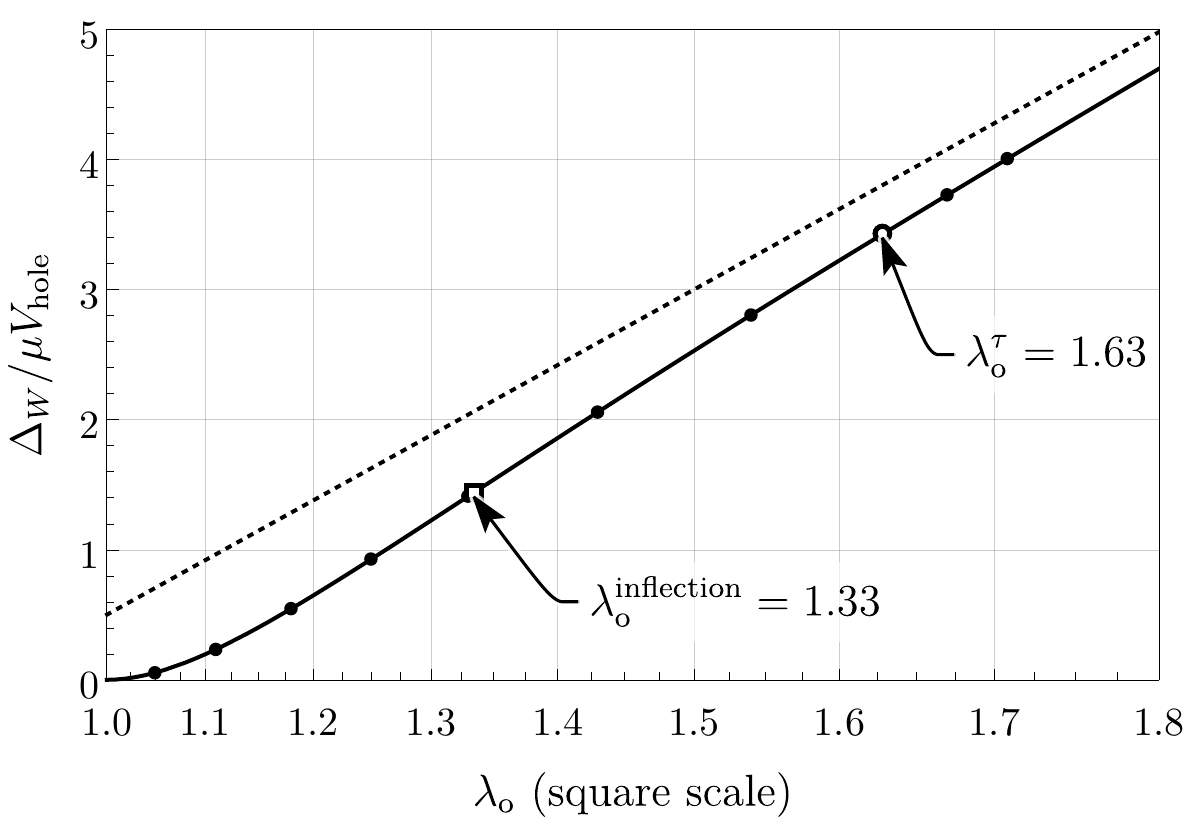}
\label{normalize by lambda_o^2}
\caption{\small{Strain energy difference normalized by $\mu V_\mathrm{hole}$ as a function of $\lambda_\mathrm{o}^2$ in the limit $R_\i/R_\o\rightarrow0$.
The continuous curve and dots correspond to the approximation and the numerical results, respectively.
The dashed curve depicts the energy difference according to the infinite deformation solution, and the square and circular marks represent the inflection and transition points.}}
\label{energy difference}
\end{figure}
Figure \ref{energy difference} shows the variations of the strain energy difference $\Delta_W$ as a function of $\lambda_\o^2$ in the limit of an infinitely small hole.
For convenience we mark the values of $\lambda_\o$ on the  horizontal axis.
Note that even tough $\Delta_W$ increases with $\lambda_\mathrm{o}$, it is rather small at small strains.
The derivative of $\Delta_W$ with respect to $\lambda_\o^2$ reaches a maximum at $\lambda_\o = 1.33 \equiv \lambda_\o^\mathrm{inflection}$, beyond which 
the slope decreases to the slope at the infinite deformation limit (the dashed line).
This finding implies that, in the case of a flawed membrane, since at small stretches $\Delta_W$ is small the likelihood for the development of a hole is small too.
 However, at larger stretches the rate at which $\Delta_W$ increases is higher than quadratic rate.
 Thus, if $\Delta_W$ is required for overcoming the surface energy for generating a hole, this result suggests that a hole will be generated only at finite strains.

\section{Conclusion}
We analyze, within the framework of finite deformation,  equvi-biaxial extension of a thin neo-Hookean circular membrane with a hole at its center.
The goal is to study how the shape and the stresses depend on the hole size and the applied stretch.
First, examining the problem in polar coordinates, we obtained the non-linear governing equation with the associated non-linear boundary conditions.
Next, by a change of variables, we introduced a simpler representation for the governing problem.
This new form lands itself to an exact closed-form solution in the limit of infinite extension.
We further propose an approximation to the solution of the problem.
Specifically, we approximate the membrane responses under intermediate and large deformations, and determine the transition stretch between them.
We reveal that at large deformations there is a need to distinguish between two regions, an outer region where the response is reminiscent of the one under infinitely stretched membrane, and an inner region in which the dependence of the stretches on the radius is weaker.
This finding maybe correlate to the wrinkling phenomenon which is frequently observed near the boundary of a hole in finitely stretched membranes.

Comparison of our solution and proposed approximations with corresponding numerical results revealed fine agreement for any hole size and applied stretch.
We find that the thickness of the membrane rapidly decreases away from the hole.
We also find that for any hole size, the hoop stress at the outer boundary is almost identical to the nominal stress in a homogeneous membrane for applied stretches larger than the transition stretch.
We reveal that in the at the surface of a small hole under an infinite deformation,
the hoop stress at is 4 times the nominal stress and the tangential stretch is twice the applied stretch.
This is twice the stress concentration factor in the small deformation limit.

Finally, we determined the strain energy stored in the punctured membrane.
As expected, we find that it is smaller than the strain energy stored in a membrane without a hole.
However, in the limit of a small hole and small deformations the difference is rather small. Only at finite deformations the difference between the energies becomes substantial, implying that a flaw in the membrane will tear out only at a finite level of stretches.

\section*{Acknowledgment}
The work was supported by the Israel Science Foundation founded by the Israel Academy of Sciences and Humanities (Grant No. 1874/16).

\begin{small}
\bibliographystyle{plainnat}
\bibliography{bibly.bib}
\end{small}
\end{document}